\documentclass[prb,twocolumn]{revtex4}

\usepackage{graphics}
\usepackage{epsfig}
\usepackage{epsf,float}
\usepackage{graphics,color}

\newcommand{\be}{\begin{equation}}
\newcommand{\ee}{\end{equation}}
\newcommand{\ben}{\begin{equation*}}
\newcommand{\een}{\end{equation*}}
\newcommand{\bea}{\begin{eqnarray}}
\newcommand{\eea}{\end{eqnarray}}
\newcommand{\bean}{\begin{eqnarray*}}
\newcommand{\eean}{\end{eqnarray*}}
\newcommand{\nn}{\nonumber}

\begin{document}
\title{Self Focussing as a Coherence Propagation Phenomena. An Application to Calculate the Coherence Length for an Atom Laser} 
\author{L. M. Castellanos\cite{lucho} and
F. E. Lopez\cite{pacho}}
\affiliation{{\small Instituto de F\'{\i}sica, Universidad de Antioquia, A.A. 1226}\\
Medell\'{\i}n-Colombia}

\begin{abstract}
A theoretical description in terms of the coherence propagation is
given for self-focussing. The concept of coherence length is defined in terms of free,  self-focussing
propagation giving results in accordance with well known experimental criteria for the laser. Extension of the method is given for an Atom Laser showing good results in agreement with recents numerical results of Trippenbach \textit {et al.} [J. Phys. B:At. Mol. Opt. Phys. {\bf{33}} 47-54 (2000)].\\   
  PACS number(s): 32.50.+d, 32.80.Qk, 32.80.-t
\end{abstract}
\maketitle

\section{introduction}

A propagating beam of particles or waves are focused in some point when they sum up constructively, and oscillate in rigorous phase at that point. The focussing of these particles or waves, is experimentaly obtained by using some optical arrangement. However, the self-focussing does not need of any experimental arrangement and raises as an unique effect no yet well explained. This became clear since the first experimental observation of self-imaging, today known as the Talbot's effect\cite{tal}.\\
In this report we formulate self-focussing as a coherence propagation phenomena and then use it as a criterion for the  definition of coherence length in both cases; interacting (Atom Laser) and non interacting (Photon Laser) particles beam propagation. \\
This paper is organized as follows: In section II we describe self-focussing as a phenomena of coherence propagation 
and obtain a master equation in Eq.(13); in section III we review the atom laser formalism and introduce the wave function $\psi_{0}$ for the condensed untrapped atoms as proposed by Gerbier \textit{et al.}\cite{Ge}, in section IV we calculate the general expression for the coherence length and finally in section V we give some relevants results and conclusions.

\section{Self-focussing of Non-interacting particles }     
The general vector state to describe a highly collimated non interacting beam
of neutral atoms propagating in a non-conducting medium is given
by
\be
   \psi(\vec{r},t)=\int_{-\infty}^{+\infty}A(\vec{k})e^{i(\vec{k}\cdot\vec{r}-\omega t )}d\vec{k}.
\ee
The equation (1) indicates a vector sum, where $\vec{k}$ corresponds to
the vector associated to the plane wave solution for free
particles. On the other hand, the integral stems from the
possibility of the particle beam of taking any (continuous)
value of momentum. In case of discrete changes in momentum it is
customary to write down:
\be
   \psi(\vec{r},t)=\sum_{k} A_{k}(\vec{k})e^{i(\vec{k}\cdot\vec{r}-\omega t )}. 
\ee

Going from Eq.(1) to Eq.(2) it is straightforward, but the former it's
not so easy to deal with. For instance, the current
density, 

\be
   \vec{J}=\frac{\hbar}{2mi}(\psi^{*}\vec{\nabla}\psi-\psi\vec{\nabla}\psi^{*}),
\ee
can be written using Eq.(2) (we drop here the time dependence)

\bea
    \vec{J}=\frac{\hbar}{2mi}[2i\sum_{k}\vec{k}|A_{k}(\vec{k})|^{2}&+& \nn\\
    \sum_{k^{'}\neq k} \sum_{k} i\vec{k}(A_{k^{'}}A^{*}_{k}e^{i(\vec{k^{'}}-\vec{k})\cdot\vec{r}}&+&
    A^{*}_{k^{'}}A_{k}e^{-i(\vec{k^{'}}-\vec{k})\cdot\vec{r}})],
\eea
 In the integral form it reads

\be
   \vec{J}=\frac{\hbar}{2m}\{2\int\vec{k}|A(\vec{k})|^{2}d\vec{k}+\int_{\vec{k'}}\Gamma(\vec{k}')d\vec{k}'\},
\ee
where

\bea
    \Gamma(\vec{k}') =\int_{k\neq k'}\vec{k}\{A(\vec{k}')A^{*}(\vec{k}) e^{-i(\vec{k}-\vec{k}')\cdot\vec{r}}+ \nn\\         A^{*}(\vec{k}')A(\vec{k})e^{i(\vec{k}-\vec{k}')\cdot\vec{r}}\}d\vec{k}.
\eea

The physical meaning of Eq.(5) is made visible when we set

\bea 
    \vec{\tau}&=&\vec{k}-\vec{k}',\nn\\ 
    d\vec{\tau}&=&d\vec{k},
\eea
for a fixed $\vec{k}'$. Then the first term at the right-hand side
of Eq.(6) (the remaining is a complex conjugate ), will be 

\bea
    \gamma&=&\int\vec{k}A(\vec{k}')A^{*}(\vec{k})e^{-i(\vec{k}-\vec{k}')\cdot\vec{r}}d\vec{k}   \nn\\
    &=&\int[A(\vec{k}-\vec{\tau})e^{i(\vec{k}-\vec{\tau})\cdot\vec{r}}]
    [\vec{k}A^{\ast}(\vec{k})e^{-i\vec{k}\cdot\vec{r}}]d\vec{k}.
\eea

Equation (8) does not quite satisfy the
correlation (classical) definition for a pair of functions $g(\xi)$
and $h(\xi)$:

\bea
    \gamma&=&g(\xi)\otimes h(\xi),\nn\\
    &=&\int g(\xi)h^{*}(\xi-\tau)d\xi,
\eea 
because in this expression $\vec{\tau}$, is a fixed correlation parameter 
and in our treatment (see Eq.(7)), $\vec{\tau}$ is changing continuously
together with $\vec{k}$. However, we still can say that Eq.(8) has the
meaning of a correlation, since the integrand can be understood as
a correlation between particles of momentum $\vec{k}$ and
$(\vec{k}-\vec{\tau})$ for a given $\vec{k}'$. Keeping this in mind
we see that equation (6) makes account of all these correlations
for diferent values of $\vec{k}'$.

Then we write:
 
\be
   \Gamma(\vec{k}')=\int[\gamma(\vec{k})+\gamma^{*}(\vec{k})]d\vec{k},
\ee
now, a particle beam can be experimentally prepared in such a way
that $A(\vec{k})$ may be a real number. This only means that for
initial condition $(t=0,r=0)$ $A(\vec{k})$ has already some definite
value (including zero). With this argument, Eq.(5) now reads: 

\bea
    &&\vec{J}=\frac{\hbar}{2m}[2\int\vec{k}|A(\vec{k})|^{2}d\vec{k}+\nn\\&&
    2\!\int\!\!d\vec{k}'\!\int\!\!
    Re[A(\vec{k}-\vec{\tau})e^{i(\vec{k}-\vec{\tau})\cdot\vec{r}}]
    [\vec{k}A^{\ast}(\vec{k})e^{-i\vec{k}\cdot\vec{r}}]d\vec{k}].
\eea

\bea
    &&\vec{J}=\frac{\hbar}{2m}\{2\int\vec{k}|A(\vec{k})|^{2}d\vec{k}+\nn\\&&
    2\int\int A(\vec{k}')A^{*}(\vec{k})\vec{k}\cos[(\vec{k}'-\vec{k})\cdot\vec{r}]d\vec{k}d\vec{k}'\}
\eea

This equation represents the more general expression for
describing a non interacting particle beam propagation. We see from here that the
second term (associated with coherence troughout
correlation), describes a focussing phenomena both directional (i.e,
certain directions contain a bigger amount of particles than other
directions) and longitudinal (along a particular direction, where
some focus will exist). In this paper we deal with the longitudinal focussing.\nn\\

\textbf{1. Longitudinal focussing phenomena}\nn\\

From the second term on the right hand side of Eq.(12) we note that
for a pair of atoms travelling in the same direction and with a
sligth difference in $\vec{k}$, the corresponding focussing along
\emph{z} will correspond to those points where:

\bea 
   (\vec{k}^{'}-\vec{k})\cdot\vec{z}=2n\pi;\\
   n=1,2,... \nn
\eea

For De Broglie particles with $\vec{p}=\hbar\vec{k}=m\vec{v}$, we have

\bea
    z\cong\frac{2n\pi \hbar}{\Delta p}=\frac{2n\pi\hbar}{m\Delta\textit{v}},
\eea
where $\Delta\textit{v}={\textit{v}}^{'}-{\textit{v}}$ is
the atomic velocity difference between atoms, $m$ the atomic mass and
$\hbar$ is the Planck constant. If the coordinate \textit{z} is
measured from the beam origin, Eq.(14) can be put in the
form

\be 
   \Delta {z}\Delta{p}\cong2n\pi \hbar, 
\ee 
to understand Eq.(15), is straightforward from the quantum mechanical
point of view. In fact we identify this expression as the
uncertainty principle since $\Delta {p}=0$ means that we can not
localize any focus on z-axis. We are in the presence of a perfect
monochromatic plane wave (i.e all every particle in the beam has exactly
the same energy).

On the other hand, if we choose any two particles into the beam
with a momentum difference $\Delta{p}$, and we track them in
time, there is a certain possibility, different from zero, that it
will focus in some point z given by Eq.(15). This probability will be
smaller as $\Delta{p}$ increases. It comes out from this
argument that, in geometrical terms, the existence of any
focus will mean some degree of coherence and therefore focus at
infinity it means perfect or total degree of coherence (for given $n$), on the
contrary focussing near the origin without any further focus, will mean a lower degree of
coherence since the beam will spread along the propagation axis.

When we are dealing with light, we better put equation (15) as:

\bea 
    (k^{'}-k)\textit{z}&=&2n\pi,  \nn\\
    \frac{1}{c}(\Delta\textit{w})\textit{z}&=&2n\pi,\nn\\ 
    \frac{2\pi\Delta\nu\Delta\textit{z}}{c}&=&2n\pi, \nn\\
    \Delta\textit{z}&=&\frac{n}{\Delta\nu}c,
\eea

For $n=1$ we obtain the so called coherence length in optics \cite{bf}
 $\Delta\textit{z}=\frac{c}{\Delta\nu}$. If we
know the band width  of a laser, we will know the distance at which, the
field will oscillate in rigorous phase. So it is interesting that
requiring focussing in Eq.(13) as a coherence condition, we
arrive at the \textit{well known} formula of coherence length.

\section{Self-focussing of Interacting particles. The Atom Laser}
Since the obtention of the first Bose Einstein Condensate (BEC) in 1995\cite{bec}, the laser of atoms became feasible.
This object is defined as a device producing an intense well collimated coherent beam of atoms\cite{kett,wis} involving a process of coherent matter-wave amplification \cite{amp}.\\
Since the atoms have masses, and they interact while travelling, the coherence of the beam presents an additional spreading, which, as variant of the Photon Laser has to be considered. We do this by replacing in Eq.(1) the outgoing wave function of the condensate $\Psi_{0}$ and then we perform the same procedure as in  section II.\\
In what follows we describe briefly the obtention of $\Psi_{0}$. A more detailed and rigorous deduction can be found  in the paper by Gerbier \textit{et al}\cite{Ge}.\\

These authors consider a BEC of  $^{87}Rb$ in the hyperfine level with $F=1$. This condensate could be in any of the sublevels $m=-1,0,1$. Atoms in a state with $m=-1$ are confined in a magnetic potential of the form $V_{trap}=\frac{1}{2}M(w^{2}_{x}x^{2}+w^{2}_{\bot}y^{2}+w^{2}_{\bot}z^{2})$; atoms in a state with $m=0$ are the untrapped ones, and those with $m=1$ are rejected out of the trap. These are the three components of the spinorial wave function of the condensate $\Psi =[\psi_m]_{m=-1,0,1}$ and they obey a set of Schr\"odinger coupled equations\cite{sh}. When the limit of weak coupling
is considered\cite{wc}, the populations $N_{m}$ satisfy $N_{1} << N_{0} << N_{-1}$ and only the states with $m=-1$, and $m=0$ are considered.\\
The condensate atoms are transferred from the state $m=-1$ (trapped) to the state $m=0$ (untrapped) by using a rf pulse 

\begin{equation}
                \vec{B}_{rf}=B_{rf}cos(\omega_{rf}t)\hat{e}_{x},	
\end{equation}
Thus the components of BEC $\psi_{m}=\psi_{m}'e^{-im\omega_{rf}t}$ satisfy the two coupled equations\cite{Ge} (after R.W.A)

\bea
	              \textit{i}\hbar\frac{\partial\psi_{-1}}{\partial t}&=&\left[\hbar\delta_{rf}+\vec{P}^{2}/2M+V_{trap}
+U\left|\psi_{-1}\right|^{2}\right]\psi_{-1}\nn\\
&&+\frac{\hbar\Omega_{rf}}{2}\psi_{0},
 \eea

\bea
 \textit{i}\hbar\frac{\partial\psi_{0}}{\partial t}&=&\left[\vec{P}^{2}/2M-Mgz
+U\left|\psi_{-1}\right|^{2}\right]\psi_{0}\nn\\
&&+\frac{\hbar\Omega_{rf}}{2}\psi_{-1},
\eea

The intensity of the interaction is given by $ U=4\pi\hbar^{2}aN/M $, where $N$ is the initial number of trapped atoms, $M$ is the atomic mass, and $a$ is the diffusion length for the interatomic collision process which, for the $^{87}Rb$ is $5nm$.\\
The uncoupling intensity between states $m=-1$, and $m=0$ is given by the Rabi flopping frequency 
\be
\hbar\Omega_{rf}=\mu_{B}B_{rf}/2\sqrt{2},
\ee 
the detunning $\delta_{rf}$ is 
\be
\hbar\delta_{rf} = V_{off}-\hbar\omega_{rf}
\ee
 and 
 \be
 V_{off}=\mu_{B}B_{0}/2+Kz^{2}/2
 \ee
 $B_{0}$ is the background magnetic field due to the coils of the trap.\\

 Equations (18) and (19) are uncoupled in the framework of the meanfield theory and the weak coupling limit \cite{Ge},  obtaining for $\psi_{0}$ , 

\be
\psi_{0}(\vec{r},t)\simeq A(\Omega_{rf},F)\frac{e^{i\frac{2}{3}\left|\zeta_{r
}\right|^{3/2}-i\frac{E_{-1}t}{\hbar}}}{\sqrt{\left|\zeta_{r}\right|^{1/2}}}
\ee
where
\be
A(\Omega_{rf},F)=-\sqrt{\pi}\frac{\hbar\Omega_{rf}}{Mgl}\phi_{-1}(x,y,z_{r})F,
\ee
here $F$ describes the finite extension of an atom laser beam due to the finite coupling time (e.g the time of rf irradiation ) and is constant for each particular laser, the adimensional parameter $ \zeta_{r}=(z-z_{r})/l $ provides a scale to the size of the trap, $ z_{r}=\eta z_{0}/2 $ is the extraction point from the trap,
and
\be
 \phi_{-1}(x,y,z_{r})=\left(\frac{\mu}{U}\right)^{1/2}[1-(x/x_{0})^{2}-(y/y_{0})^{2}-(z/z_{0})^{2}]^{1/2}
\ee
The quantity $\mid \phi_{-1}(x,y,z_{r})\mid^2$, corresponds to the trapped atomic population in the ouput point $z_{r}$.\\ 
In what follows we define some important figures such as $l$, $z_{0}$, $x_{0}$, $y_{0}$ and $\eta$
 \bea
 l&=&\left(\frac{\hbar^2}{2M^2g}\right)^{1/3}\nn\\
 x_{0}^{2}&=&2\mu/Mw^{2}_{x},\nn\\
  y_{0}^{2}&=&2\mu/Mw^{2}_{\bot},\nn\\
  z_{0}^{2}&=&2\mu/Mw^{2}_{\bot},\nn\\
  \eta &=&(2\hbar\delta_{rf}+4\mu/7)\frac{1}{2mgz_{0}}
 \eea
  where $\mu$ is the chemical potential, and is understood as the neccessary energy to either $\textit{add or remove}$  a condensed atom in the trap ensemble. The chemical potential
  $\mu$ is defined\cite{peth} 
  \be
  \mu =\left(\frac{\hbar\omega}{2}\right)\left(\frac{15aN_{-1}}{\sigma}\right)^{2/5}
  \ee 
  with  $\omega =(\omega_{x}\omega^2_{\bot})^{1/3}$ and the harmonic oscillator length defined as  $\sigma = (\hbar/M\omega)^{1/2}$, note that $l<<x_{0},y_{0},z_{0}$ and for $^{87}Rb$ it is known that  $l\simeq 0.28\mu m$

\section{Coherence Length For an Atom Laser}
In the spirit of section II, we now use Eq.(13) in order to find the coherence length. To do this we need to know the propagation vector ${\vec{k}}$, which is calculated using Eq.(23) and following the standard procedures (see for example Fl\"ugge\cite{flu}).\\
Along these lines we then calculate $\vec{J}$, and find $\vec{v} = \vec{J}/\rho$ with $\rho = \mid\psi_{0}\mid^2$
\bea
\vec{J}&=&\frac{\hbar}{2mi}(\psi_{0}^{*}\vec{\nabla}\psi_{0}-\psi_{0}\vec{\nabla}\psi_{0}^{*})\nn\\
v&=&\frac{{J}}{\rho}=\frac{\hbar}{ml}\sqrt{\zeta_{r}}\nn\\
v&=&\frac{\hbar}{m}\frac{\sqrt{z+z_{r}}}{l^{3/2}}
\eea
here $\zeta_{r}=\frac{z}{l}+\frac{z_{r}}{l}$. We now use the De Broglie relation $\vec{p}=\hbar {k} = m\vec{v}$ to obtain  the propagation vector $\vec{k}$, we obtain for its magnitude 
\be
k = \frac{\sqrt{z+z_{r}}}{l^{3/2}}
\ee

We then replace Eq.(29) into Eq.(13)
\be
((z+z_{r})^{1/2}-(z^{\prime}+z_{r})^{1/2})z=2n\pi l^{3/2},
\ee
It is clear from section II and Eq(13) that $z-z^{\prime}$ is the correlation length for the interacting atom laser beam between two diferent points of the beam. Since we are interested in the coherence length measured from the extraction point $z_{r}$, we make $z^{\prime}= z_{r}$. therefore 
\be
((z+z_{r})^{1/2}-(2z_{r})^{1/2})z=2n\pi l^{3/2},
\ee
 By solving this equation  for $n=1$,  we obtain the coherence length $z$ for the atom laser
\section {Numerical Results and Conclusions} 
 
 We use the mathematica program to solve numericaly Eq.(31), below we list the results for all possible atoms laser with alcalin species 

\begin{eqnarray}
	^{23}Na&=&2.4622\mu m, \nn\\
	^{87}Rb&=&1.0299\mu m,  \nn\\
	^{7}Li&=&5.4461\mu m. \nn
\end{eqnarray}
In particular, our treatment leads to a quantitative agreement with the experimental results of Marek Trippenbach \textit{et al.}\cite{Tri} for atoms of $^{23}Na$. In their experiment they obtain diferent results for diferents $w_{rf}$, this results  running between 2.0 and 5.0 $\mu m$.\\

In conclusion, we have proposed a very simple form for the calculation of the coherence length for an Atom Laser, and shown the validity for the method by comparison with the numerical results of Trippenbach et al. As a final coment we note tha the coherence length is not depending on the iradiation time

\end{document}